\def\nk{n_{\rm b}}

\def\rfr#1{Equation~(\ref{#1})}
\def\rfrs#1#2{Equations~(\ref{#1})~to~(\ref{#2})}

\def\derp#1#2{\rp{\partial{#1}}{\partial{#2}}}
\def\dert#1#2{\frac{{{\textrm{d}}}{#1}}{{{\textrm{d}}}{#2}}}
\def\virg#1{``#1"}

\def\eqi{\begin{equation}}
\def\eqf{\end{equation}}
\def\eqia{\begin{eqnarray}}
\def\eqfa{\end{eqnarray}}

\def\rp#1#2{{#1\over#2}}
\def\lb#1{\label{#1}}

\def\bds#1{\boldsymbol{#1}}
\def\cO{\cos\Omega}
\def\sO{\sin\Omega}

\def\cI{\cos I}
\def\sI{\sin I}

\def\ton#1{\left(#1\right)}
\def\qua#1{\left[#1\right]}
\def\grf#1{\left\{#1\right\}}

\documentclass[onecolumn]{aastex}

\usepackage[title]{appendix}
\usepackage{hyperref}
\usepackage{booktabs}
\usepackage[table,xcdraw]{xcolor}
\usepackage{multirow}
\usepackage{rotating,tabularx}
\usepackage{float}
\usepackage{amsmath,textgreek,w-greek,wasysym}
\usepackage{amsthm}
\usepackage{amscd,lineno}
\usepackage{amssymb,dsfont}
\usepackage{graphicx,epsfig}
\usepackage{txfonts}
\usepackage{xr-hyper}
\RequirePackage{color}

\newcommand{\emaila}{lorenzo.iorio@libero.it}

\allowdisplaybreaks[1]

\begin{document}

\title{On testing frame-dragging with LAGEOS and a recently announced geodetic satellite}

\shortauthors{L. Iorio}

\author{Lorenzo Iorio\altaffilmark{1} }
\affil{Ministero dell'Istruzione, dell'Universit\`{a} e della Ricerca
(M.I.U.R.)-Istruzione
\\ Permanent address for correspondence: Viale Unit\`{a} di Italia 68, 70125, Bari (BA),
Italy}

\email{\emaila}

\begin{abstract}
Recently, Ciufolini and coworkers announced the forthcoming launch of a new cannonball geodetic satellite in 2019. It should be injected in an essentially circular path with the same semimajor axis $a$ of LAGEOS, in orbit since 1976, and an inclination $I$ of its orbital plane supplementary with respect to that of its existing cousin. According to their proponents, the sum of the satellites' precessions of the longitudes of the ascending nodes $\Omega$ should allow \textcolor{black}{one} to test the general relativistic Lense-Thirring effect to a $\simeq 0.2\%$ accuracy level, with a contribution of the mismode\textcolor{black}{l}ing in the even zonal harmonics $J_\ell,~\ell=2,4,6,\ldots$ of the geopotential to the total error budget as little as $0.1\%$. Actually, such an ambitious goal seems to be hardly attainable because of the direct and indirect impact of, at least, the first even zonal $J_2$. On the one hand, the lingering scatter of the est\textcolor{black}{i}mated values of such a key geophysical parameter from different recent GRACE/GOCE-based global gravity field solutions is representative of an uncertainty which may directly impact the summed Lense-Thirring  node precessions at a $\simeq 70-80\%$ in the worst scenarios, and to a $\simeq 3-10\%$ level in other, more favorable cases. On the other hand, the phenomenologically measured secular decay $\dot a$ of the semimajor axis of LAGEOS (and, presumably, of the other satellite as well), currently known at a $\sigma_{\dot a}\simeq 0.03~\textrm{m~yr}^{-1}$ level after more than 30 yr, will couple with the sum of the $J_2$-induced node precessions yielding an overall bias as large as $\simeq 20-40\%$ after $5-10$ yr. A further systematic error of the order of $\simeq 2-14\%$ may arise from an analogous interplay of the secular decay of the inclination $\dot I$ with the oblateness-driven node precessions.
\end{abstract}

keywords{
Experimental studies of gravity; Experimental tests of gravitational theories; Satellite orbits; Harmonics of the gravity potential field
}

%keywords{Astrophysical studies of gravity; General relativity; Cosmological constant; Neutron stars \& pulsars; Classical black holes}
%

\section{Introduction}
The cannonball geodetic satellites of the LAGEOS family, i.e. LAGEOS (L), LAGEOS II (L II) and LARES (LR), entirely covered by passive retroreflectors and tracked on a continuous basis from several ground stations scattered around the world with the Satellite Laser Ranging (SLR) technique \citep{Combrinck2010}, are currently used, among other things, to put to the test some predictions of the Einstein's General Theory of Relativity (GTR) \citep{2011SAJG..114..549C,Combrinck2013,2014AdHEP2014E...1P,2015CQGra..32o5012L,7119152}. One of them is known as\footnote{Recent historical studies have pointed out that it would be more correct to name it as Einstein-Thirring-Lense effect \citep{2007GReGr..39.1735P,2008mgm..conf.2456P,Pfister2014}. Nonetheless, we will follow the denomination now commonly used.} Lense-Thirring (LT) effect \citep{LT18}, and its measurement is one of the current goals of the LAGEOS-type satellites in fundamental physics. It consists of small secular precessions of some orbital elements of a test particle in geodesic motion around a rotating primary which, in the case of the aforementioned spacecraft, amount to some dozens milliarcseconds per year (mas yr$^{-1}$). Such long-term orbital rates of change are induced by the post-Newtonian (pN) gravitomagnetic field \citep{Thorne86,1988nznf.conf..573T} of the central body. It is generated by the mass-energy currents of its angular momentum $\bds S$, and it is believed to play an important role in several dynamical effects taking place around spinning Kerr black holes \citep{
1971NPhS..229..177P,1988nznf.conf..573T,2005NYASA1045..232W}. For an overview of the so-far performed attempts to measure it with artificial satellites in the field of the Earth, see, e.g., \citet{2011Ap&SS.331..351I,2013NuPhS.243..180C,2013CEJPh..11..531R}, and references therein.

The space environment of an astronomical body like the Earth is characterized by several other competing accelerations, both of non-gravitational \citep{Sehnal1975,1987ahl..book.....M,1994AdSpR..14...45D} and gravitational \citep{1974RvGSP..12..421L,Rox86,Kaula00} origin, some of which have just the same temporal signature of the relativistic ones of interest. \textcolor{black}{In view of their relatively small magnitudes with respect to the Newtonian gravitational monopole of the Earth, they can be treated perturbatively with the standard methods of perturbation theory and celestial mechanics; see, e.g., \citet{2013osos.book.....X,Bertotti03,2011rcms.book.....K,2014grav.book.....P}.} All of them \textcolor{black}{contribute} in determining the actual path followed by a test particle which, thus, deviates more or less notably from the Keplerian ellipse \citep{2014hso..book.....C}.  As a consequence,  extracting the  gravitomagnetic signature  and assessing realistically the uncertainty with which such a task can be implemented is not easy, and requires a careful analysis of all such other biasing disturbances. \textcolor{black}{Since the gravitomagnetic effects are linear trends (see \rfr{LTO} below), this is particularly true for those perturbations which are either secular rates too, like those due to the asphericity of the Earth's gravitational potential \citep{Rox86,Kaula00} (see \rfr{NodoJ2} below) or to some non-gravitational accelerations \citep{2002P&SS...50.1067L}, or are harmonic with so small characteristic frequencies that they may mimic the action of biasing linear trends over observational time spans which cover just a fraction of their period of variation like certain tidal perturbations \citep{2002IJMPD..11..599P}.}

In 1976, \citet{1976CeMec..13..429V,1976PhRvL..36..629V} noticed that the nodes $\Omega$ of two counter-revolving satellites sharing the same orbital parameters undergo identical secular LT precessions
\eqi
\dot\Omega_\textrm{LT} = \rp{2GS}{c^2 a^3\ton{1-e^2}^{3/2}}\lb{LTO}
\eqf
and opposite secular Newtonian rates of change due to the even zonal harmonic coefficients $J_\ell,~\ell=2,4,6\ldots$ of the multipolar expansion of the Earth's gravitational potential. \textcolor{black}{In \rfr{LTO}, $G,~c$ are the Newtonian constant of gravitation and the speed of light in vacuum, respectively, while $a,~e$ are the semimajor axis and the eccentricity of the satellite's orbit, respectively.} The largest one is induced by the first even zonal harmonic $J_2$: it is
\eqi
\dot\Omega_{J_2} = -\rp{3}{2}\nk\ton{\rp{R}{a}}^2\rp{\cos I J_2}{\ton{1-e^2}^2},\lb{NodoJ2}
\eqf
and its nominal value is usually several orders of magnitude larger than the gravitomagnetic one of \rfr{LTO}; see Table~\ref{tavola0} for the currently orbiting satellites of the LAGEOS family, their relevant orbital parameters and precessions. \textcolor{black}{In \rfr{NodoJ2}, $R$ is the Earth's mean equatorial radius, while $\nk,~I$ are the Keplerian mean motion and the inclination to the Earth's equator of the satellite's orbit, respectively. } At that time, the state-of-the-art of modeling the Earth's geopotential would \textcolor{black}{not have allowed one} to use the node of a single satellite because of the still too large uncertainty in $J_2$. It would have yielded a mismodelled node precession which would have completely overwhelmed the LT one in view of its much larger size and identical temporal pattern. Such a state of affairs still persists today, despite the steady efforts in producing global gravity field models of ever increasing accuracy by several institutions throughout the world. \textcolor{black}{Unfortunately, there are no other Keplerian orbital elements affected by both the gravitomagnetic field of the Earth and the geopotential with different temporal patterns, so that they could be used to decouple their mutual impact. Indeed, only the perigee  undergoes a secular Lense-Thirring rate; actually, apart from the fact that it is much more heavily impacted by the non-gravitational perturbations than the node, also the even zonal harmonics $J_\ell,~\ell=2,4,6\ldots$ induce just the same kind of linear trend on it \citep{Kaula00}. }
Thus, \citet{1976CeMec..13..429V,1976PhRvL..36..629V} considered the sum of the nodes of both their counter-orbiting spacecraft, which should have been endowed with drag-free apparatus to counteract the non-gravitational perturbations. Indeed, such an arrangement would allow, at least in principle, to exactly cancel out the  classical perturbations due to the even zonals and add up the LT ones. In 1986, \citet{CiufoPRL86} put forth an essentially equivalent version of the scenario by \citet{1976CeMec..13..429V,1976PhRvL..36..629V} suggesting to launch a new passive geodetic satellite $X$ with the same orbital configuration of LAGEOS, launched in 1976, apart from the inclination $I_X$ which should have been displaced by 180 deg from that of the already existing spacecraft.
%See Table~\ref{tavola0} for the orbital parameters of LAGEOS.
Such a \virg{butterfly} orbital geometry has the same main features of that by \citet{1976CeMec..13..429V,1976PhRvL..36..629V} as far as the classical and relativistic node precessions are concerned. Thus, also \citet{CiufoPRL86}  proposed to monitor the sum of the nodes of LAGEOS and of the proposed new spacecraft $X$. The non-gravitational perturbations affecting the nodes would not have posed a too severe threatening to the LT measurement because of their relatively accurate modeling for cannonball satellites like the LAGEOS-type ones\textcolor{black}{, being at the $\lesssim 1\%$ level of the LT}. Such a conclusion is still valid today\textcolor{black}{; see, e.g., \citet{1981CeMec..25..169S,2001P&SS...49..447L,2002P&SS...50.1067L,2015CQGra..32o5012L,2016MNRAS.460..802L,2017AcAau.140..469P}, and references therein}.

In the following years, LAGEOS II and LARES were actually launched,
%(see Table~\ref{tavola0} for their orbital parameters),
but none of them in the orbital configuration desired by \citet{CiufoPRL86}  for his satellite which assumed various provisional names over the years like Lageos-3, LARES/WEBER-SAT. Various LT tests conducted by combining data of LAGEOS and LAGEOS II \citep{2013NuPhS.243..180C}, and more recently also of LARES \citep{2016EPJC...76..120C}, have been reported so far by Ciufolini and collaborators; there is currently a lingering debate on several issues which would plague them like their realistic overall accuracy \citep{2011Ap&SS.331..351I,2013NuPhS.243..180C,2013CEJPh..11..531R}.

Recently, \citet{2017EPJP..132..336C} announced that a further LAGEOS-type satellite, which we will denote as CiufoLares (CL) in honor of \textcolor{black}{its} proponent instead of the rather anodyne LARES 2, is planned for launch in 2019 with the new VEGA C rocket. This time, the orbit of the forthcoming SLR target seems to be the right one: indeed, it should match the originally proposed butterfly configuration with LAGEOS, up to allowed offsets in $a$ and $I$ as little as $\Delta a^\textrm{CL}=20~\textrm{km},~\Delta I^\textrm{CL}=0.15~\textrm{deg}$. \citet{2017EPJP..132..336C} claimed an overall accuracy in the LT measurement of the order of $0.2\%$, with a total contribution from the uncertainties in the even zonal harmonics as little as $\simeq 0.1\%$.

In our preliminary analysis, we will show that, actually, the overall impact of just the first even zonal harmonic of the geopotential, including both its direct effect due to the mismode\textcolor{black}{l}ing in $J_2$ and the indirect one due to the interplay with the measured secular decay of the semimajor axis, and, perhaps, of the inclination as well, of LAGEOS and, likely, of CL as well, may be up to $200-800$ times larger over a data analysis $5-10$ yr long. The paper is organized as follows. In Section~\ref{geopot}, we will deal with the currently existing scatter in the estimated values of the Earth's oblateness from the latest global gravity field solutions produced by several independent institutions. Instead of  taking into account the more or less realistically re-scaled sigmas of just a single, favorable Earth's gravity field, which is also likely a-priori imprinted by the LT itself, by comparing the estimated values of the coefficient ${\overline{C}}_{2,0}$ of several pairs global gravity models of similar accuracy, we will show that the total bias in the LT summed node rates can be as large as up to $\simeq 10-80\%$ for certain  geopotential models. \textcolor{black}{Here, ${\overline{C}}_{2,0}$ is the fully normalized Stokes coefficient of degree $\ell = 2$ and order $m = 0$ of the  multipolar expansion of Earth's gravitational potential \citep{2010ITN....36....1P}. It is related to the first even zonal harmonic of the geopotential by $J_2 = -\sqrt{5}~{\overline{C}}_{2,0}$.} Section~\ref{decay} is devoted to quantitatively assessing the indirect effect of the secular orbital decays measured so far for all the existing satellites of the LAGEOS family on their node precessions due to $J_2$ showing that the resulting systematic bias can reach $\simeq 20-40\%$ of the combined LT node signal after $5-10$ yr. An analogous effect due to a possible secular decay of the inclination, yielding a bias of the order of $\simeq 2-14\%$, is dealt with in Section~\ref{decayI}. Section~\ref{finale} summarizes our findings and offers our conclusions. \textcolor{black}{A list of definitions and conventions used throughout the text is displayed in Appendix~\ref{appen} for the benefit of the reader. Appendix~\ref{tavole} contains the tables and the figures.}
\section{The direct impact of the mismodeling of the even zonal harmonics of the geopotential}\lb{geopot}
By introducing the coefficient
\eqi
\dot\Omega_{.2} =  -\rp{3}{2}\nk\ton{\rp{R}{a}}^2\rp{\cos I}{\ton{1-e^2}^2},\lb{OJ2}
\eqf
the mismodelled part of the sum of the node precessions of L and CL due to our imperfect knowledge of the Earth's oblateness can be calculated as
\eqi
\delta\ton{\dot\Omega_{J_2}^\textrm{tot}}\leq \left|\dot\Omega^\textrm{L}_{.2}+\dot\Omega^\textrm{CL}_{.2}\right|\delta J_2,\lb{sumerrJ2}
\eqf
where $\delta J_2$ represents some quantitative measure of the actual uncertainty in the first even zonal harmonic of the geopotential.
Note that, since $J_2$ is an overall external parameter not depending on the satellite, the orbital configuration of CL is crucial in the assessment of \rfr{sumerrJ2} since the coefficient proportional to $\delta J_2$ is made just of the sum of the $J_2-$induced node rates of L and CL. \textcolor{black}{In principle, if the orbital parameters of CL were exactly equal to their nominal values, \rfr{sumerrJ2} would vanish since $\dot\Omega^\textrm{L}_{.2} = -\dot\Omega^\textrm{CL}_{.2}$}.
The percent error $\psi$ in the sum of the Lense-Thirring node rates can be straightforwardly evaluated as the ratio of \rfr{sumerrJ2} to the sum of the nominal gravitomagnetic node precessions
\eqi
\dot\Omega_\textrm{LT}^\textrm{tot}=\dot\Omega_\textrm{LT}^\textrm{L} + \dot\Omega_\textrm{LT}^\textrm{CL},
\eqf
i.e.
\eqi
\psi\doteq\rp{\delta\ton{\dot\Omega^\textrm{tot}_{J_2}}}{\dot\Omega_\textrm{LT}^\textrm{tot}}\lb{psi}.
\eqf
The key factors in \rfr{sumerrJ2} will be the actual departures $\Delta a^\textrm{CL},~\Delta I^\textrm{CL}$ of the CL orbit with respect to the ideal one, and a realistic evaluation of the lingering uncertainty in $J_2$. About the orbital configuration of CL, it should be noted that the inclination of the existing LARES satellite, which should have been inserted in an orbital plane supplementary to that of LAGEOS, exhibits an offset of $\Delta I^\textrm{LR}=0.7~\textrm{deg}$ with respect to the ideal case; thus, in the following we will prudentially assume for CL  $\Delta I^\textrm{CL}=0.5~\textrm{deg}$ with respect to the smaller value quoted in Table~\ref{tavola0}.

As far as the evaluation of the uncertainty in $J_2$ is concerned, an ever increasing number of global gravity field solutions produced by several independent institutions throughout the world \textcolor{black}{chiefly} from data of the dedicated GRACE and GOCE space-based missions is nowadays available. They are made freely available by the International Center for Global Earth Models (ICGEM) which collects them in its webpage at http://icgem.gfz-potsdam.de/tom$\_$longtime on the Internet. Thus, it is neither realistic to assume the mere statistical, formal errors $\sigma_{J_2}$ released by the various models as a trustable measure of $\delta J_2$  nor to pick up just that model which gives the smallest overall uncertainty in the LT test by discarding other ones if less favorable. Furthermore, it must also be stressed that some models return determinations of ${\overline{C}}_{2,0}$ obtained \textcolor{black}{directly} from the constellation of the existing SLR satellites among which the LAGEOS family plays an important role; such values  must be, of course, discarded to avoid any possible a priori \virg{imprint} of GTR itself
\citep[pag. 1718]{1996NCimA.109.1709C}. Finally, in order to meaningfully compare the values of the first even zonal retrieved from different models, it is important that they share the same tide system (zero-tide or tide-free). \textcolor{black}{For an explanation of such definitions, see
\citet[Sect.~1.1~and~6.2.2]{2010ITN....36....1P}.}
Actually, \citet{2017EPJP..132..337C} did not take ito account any of such issues. Indeed, instead of considering several different global gravity field solutions, they came to their claimed $0.1\%$ error due to the even zonals by using only a single Earth's gravity model, i.e. GOCO05s \citep{goco05s}, by using its $1-\sigma$ formal errors, apart from ${\overline{C}}_{2,0}$ whose uncertainty was specifically evaluated in a quite hand-waving, confusing and arbitrary fashion. Suffice it to say that \citet{2017EPJP..132..337C} resorted to a historical time series of its SLR-based determinations covering a temporal interval (1975-2010) which will necessarily have nothing to do with L and CL. As a further drawback, such a model relies upon data records from CHAMP, GRACE, GOCE and six SLR satellites (LAGEOS, LAGEOS II, Starlette, Stella, Ajisai, Larets). The data from the geodetic satellites are routinely used in some global gravity field models to determine more accurately just the even zonals of low degree. Thus, the values for ${\overline{C}}_{2,0}$ from GOCO05s are unavoidably plagued by an a-priori \virg{imprint} of the LT effect itself which mainly concentrates just in the even zonals of the lowest degree \citep[pag. 1719]{1996NCimA.109.1709C}. As such, GOCO05s should not be considered as a viable background model to be used in dedicated data reductions to measure the gravitomagnetic field involving LAGEOS itself.
Furthermore, also the evaluation of the error in ${\overline{C}}_{2,0}$ proposed by \citet{2017EPJP..132..337C} should be deemed as a priori strongly imprinted by the LT itself since it entirely relies upon SLR data.
%\textcolor{black}{Interestingly, \citet{2017EPJP..132..337C} omitted to mention the role of the SLR data of the geodetic satellites in GOCO05s, a fact which is explicitly stated even in the webpage of the Gravity Observation Combination (GOCO) consortium  at https://www.bgu.tum.de/iapg/forschung/schwerefeld/goco/. Indeed,  \citet[pag.~2]{2017EPJP..132..337C} explicitly wrote about such a model: \virg{It is estimated from data of the satellite gravity missions GOCE, GRACE and CHAMP}.}
It would have been much more useful and appropriate if \citet{2017EPJP..132..337C} had simulated a realistic data reduction of LAGEOS and the new satellite by simultaneously estimating, among other things,  both a LT parameter, or, even better, the Earth's angular momentum $\bds S$ itself, and ${\overline{C}}_{2,0}$ on some predetermined temporal basis (say, weekly, monthly, etc.), and the resulting correlations between $\bds S$ and ${\overline{C}}_{2,0}$  had been inspected. Such a procedure should have been repeated for several different Earth's global gravity field solutions not including previous SLR data from LAGEOS itself as background models. Inexplicably, it has never been yet implemented, at least publicly, by any group,  not even with the real data of the existing  satellites of the LAGEOS family, after more than 20 yr since the first published tests \citep{1996NCimA.109..575C}.
\textcolor{black}{About the issue of the a priori \virg{imprint} of the LT in the existing estimated values of ${\overline{C}}_{2,0}$, it should be noted that, although indirectly, it may somewhat affect also some models based solely on GRACE/GOCE data. Indeed, many of them use previously obtained global solutions which rely upon just SLR data for the low-degree even zonals as background gravity models.}

\begin{figure*}
\centerline{
\vbox{
\begin{tabular}{c}
\epsfbox{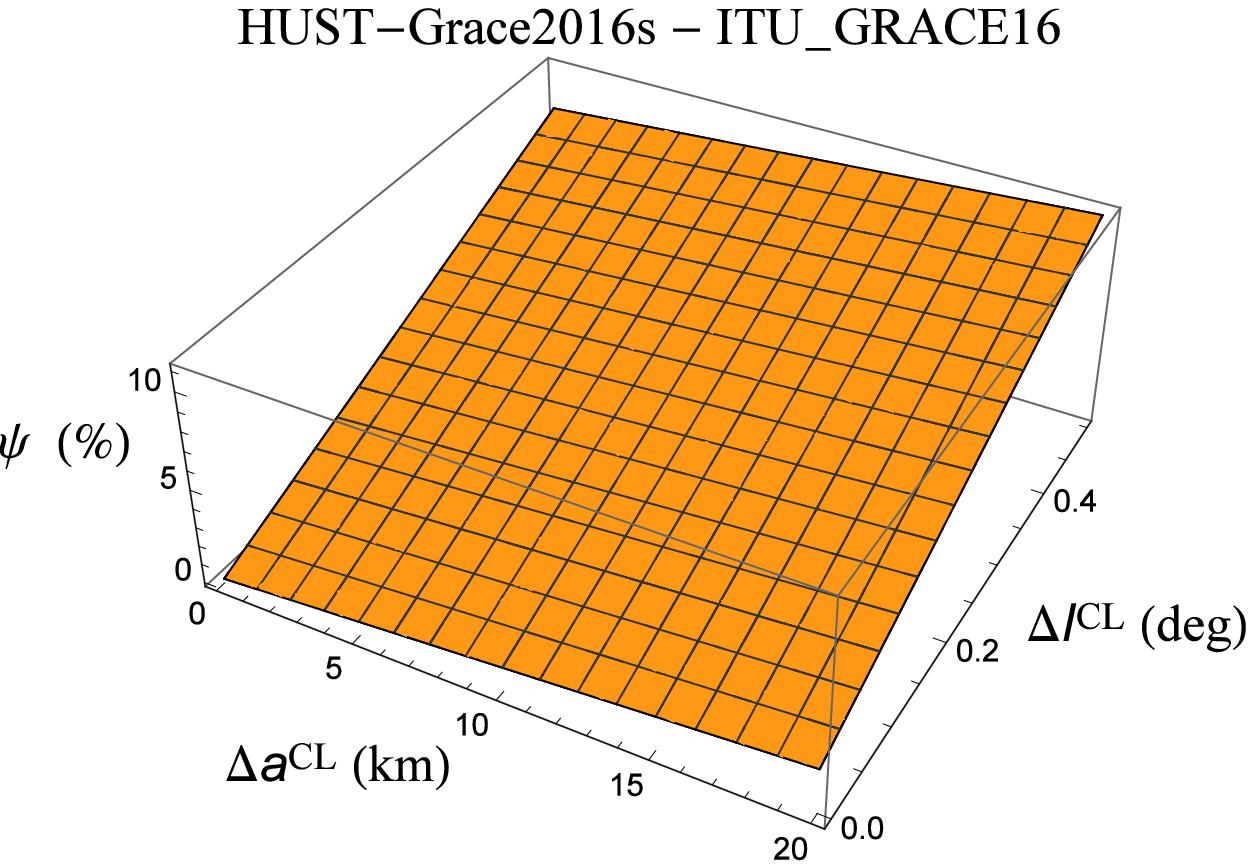}\\
\end{tabular}
}
}
\caption{Percent systematic error $\psi$ in the total $\textrm{L} + \textrm{CL}$ Lense-Thirring node shift due to the mismodeling in $J_2$ calculated with \rfr{psi} as a function of $\Delta a^\textrm{CL},~\Delta I^\textrm{CL}$. The absolute value of the difference $\Delta{\overline{C}}_{2,0}$ between the estimated values of ${\overline{C}}_{2,0}$ of the HUST-Grace2016s \textcolor{black}{\citep{2017AdSpR..60..597Z}} and ITU$\_$GRACE16 \textcolor{black}{\citep{ITU16}} zero-tide models is assumed as representative of the uncertainty $\delta{\overline{C}}_{2,0}$. Cfr. with Table~\ref{zerotide}.}\label{figzerotide}
\end{figure*}

In Tables~\ref{zerotide}~to~\ref{tidefree}, we adopt the differences $\Delta {\overline{C}}_{2,0}$ between the nominal values of the estimated coefficients ${\overline{C}}^{i/j}_{2.0}$ of several pairs of geopotential models $i,~j$\textcolor{black}{, not directly relying upon the SLR data of the LAGEOS family itself,} for the uncertainty $\delta J_2$  entering \rfr{sumerrJ2}; \citep[e.g.][pag. 1713]{1996NCimA.109.1709C}.
From Table~\ref{zerotide}, which compares some models of the zero-tide system \textcolor{black}{built from GRACE data records differing by their temporal extensions and type of measurements}, it turns out that the maximum uncertainty in the systematic bias due to the first even zonal can reach the $\simeq 10\%$ level by using the GRACE-based HUST-Grace2016s \textcolor{black}{\citep{2017AdSpR..60..597Z}} and ITU$\_$GRACE16 \textcolor{black}{\citep{ITU16}} models; cfr. with Figure~\ref{figzerotide}.
Interestingly, the formal errors of such solutions are comparable \citep[see][pag. 91]{2009SSRv..148...71C}; indeed, it is
\eqi
\sigma^\textrm{HUST-Grace2016s}_{{\overline{C}}_{2,0}}=9.6\times 10^{-14},~\sigma^\textrm{ITU$\_$GRACE16}_{{\overline{C}}_{2,0}}=1.2\times 10^{-13},
\eqf
so that
\eqi
\rp{\sigma^\textrm{ITU$\_$GRACE16}_{{\overline{C}}_{2,0}}}{\sigma^\textrm{HUST-Grace2016s}_{{\overline{C}}_{2,0}}} =1.25.
\eqf
\textcolor{black}{HUST-Grace2016s is a static global gravity field model obtained by using approximately 13 yr (spanning from January 2003 to April 2015) of GRACE-only data \citep{2017AdSpR..60..597Z}. ITU$\_$GRACE16 is a static global gravity field model computed from GRACE Satellite-to-Satellite-Tracking (SST) data of 50 months collected between April 2009 to October 2013 \citep{ITU16}. For the details of the other models used in Table~\ref{zerotide}, see the dedicated entries at the ICGEM Webpage http://icgem.gfz-potsdam.de/tom$\_$longtime.}

\begin{figure*}
\centerline{
\vbox{
\begin{tabular}{c}
\epsfbox{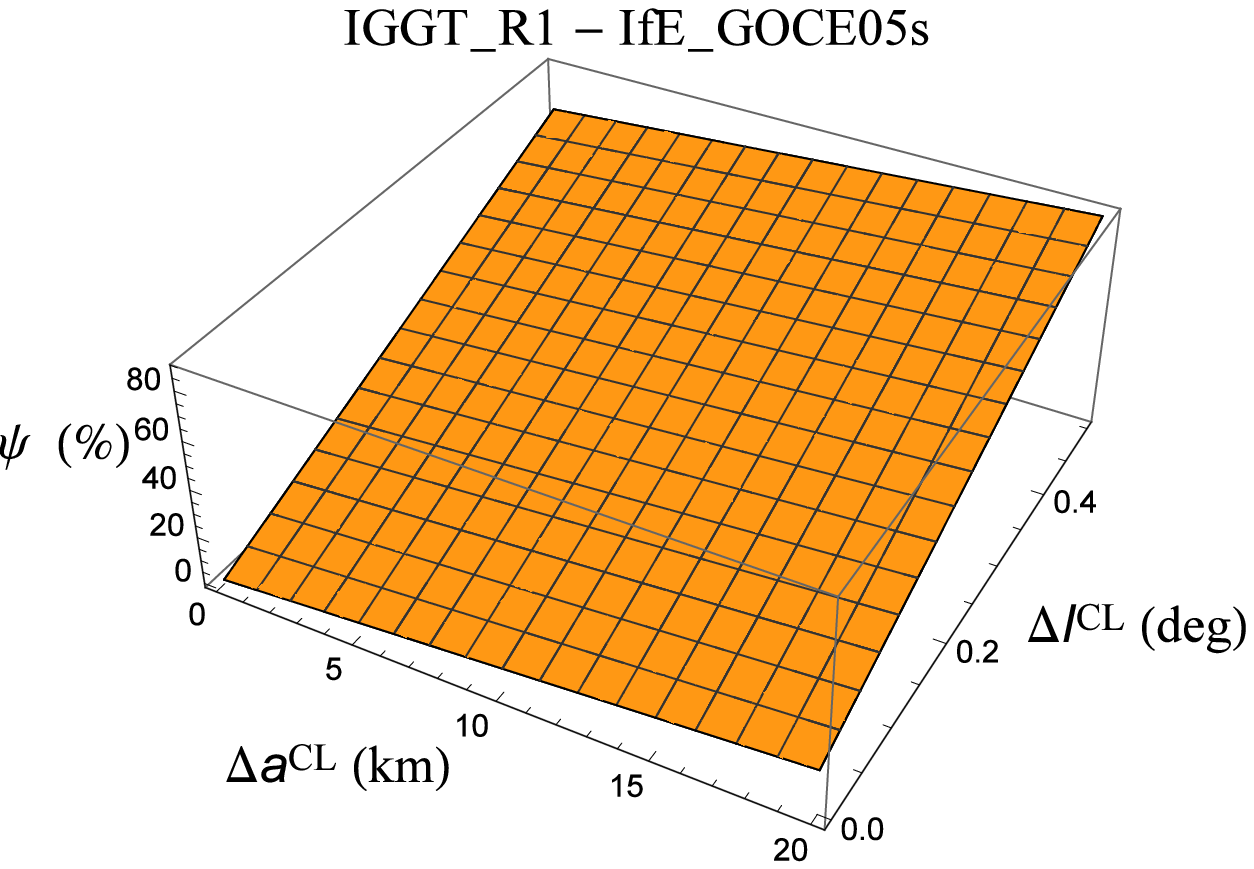}\\
\end{tabular}
}
}
\caption{Percent systematic error $\psi$ in the total $\textrm{L} + \textrm{CL}$ Lense-Thirring node shift due to the mismodeling in $J_2$ calculated with \rfr{psi} as a function of $\Delta a^\textrm{CL},~\Delta I^\textrm{CL}$. The absolute value of the difference $\Delta{\overline{C}}_{2,0}$ between the estimated values of ${\overline{C}}_{2,0}$ of the IGGT$\_$R1 \textcolor{black}{\citep{2018JGeod..92..561L}} and IfE$\_$GOCE05s \textcolor{black}{\citep{IFE05}} tide-free models is assumed as representative of the uncertainty $\delta{\overline{C}}_{2,0}$. Cfr. with Table~\ref{tidefree}.}\label{figtidefree}
\end{figure*}

Table~\ref{tidefree}, showing the impact of some tide-free models \textcolor{black}{from various GOCE data records analyzed with different approaches}, tells us that the $J_2$-driven overall uncertainty in the LT test can be as large as $\simeq 85\%$ if the GOCE-based IGGT$\_$R1 \textcolor{black}{\citep{2018JGeod..92..561L}} and IfE$\_$GOCE05s \textcolor{black}{\citep{IFE05}} models are considered; see also Figure~\ref{figtidefree}. Moreover, the (formal) released sigmas of IGGT$\_$R1 and IfE$\_$GOCE05s are comparable since
\eqi
\sigma^\textrm{IGGT$\_$R1}_{{\overline{C}}_{2,0}}=9.7\times 10^{-11},~\sigma^\textrm{IfE$\_$GOCE05s}_{{\overline{C}}_{2,0}}=1.8\times 10^{-11},
\eqf
so that
\eqi
\rp{\sigma^\textrm{IGGT$\_$R1}_{{\overline{C}}_{2,0}}}{\sigma^\textrm{IfE$\_$GOCE05s}_{{\overline{C}}_{2,0}}} = 5.4.
\eqf
\textcolor{black}{IGGT$\_$R1 \citep{2018JGeod..92..561L} relies upom the use of the  three invariants of the gravitational gradient tensor (IGGT) to process about 1 yr of GOCE data (2009-2010), while in IfE$\_$GOCE05s \citep{IFE05} almost the same GOCE-only data set is analyzed with either the SST and the Satellite Gravity Gradient (SGG) techniques. For the details of the other tide-free models used in Table~\ref{tidefree}, see their dedicated entries at http://icgem.gfz-potsdam.de/tom$\_$longtime.}

\textcolor{black}{It is difficult to properly argue about the differences between Table~\ref{zerotide} and Table~\ref{tidefree}, also because, after all, new global gravity field solutions will be available if and when the LT-dedicated  L-CL analysis will be performed. Our results should likely be looked just as illustrative of the current state-of-affairs and of the need of not limiting just to one particular model, given the sound possibility that the increasing number of values of $J_2$ will not finally converge to the desired level of accuracy.}

\begin{figure*}
\centerline{
\vbox{
\begin{tabular}{c}
\epsfbox{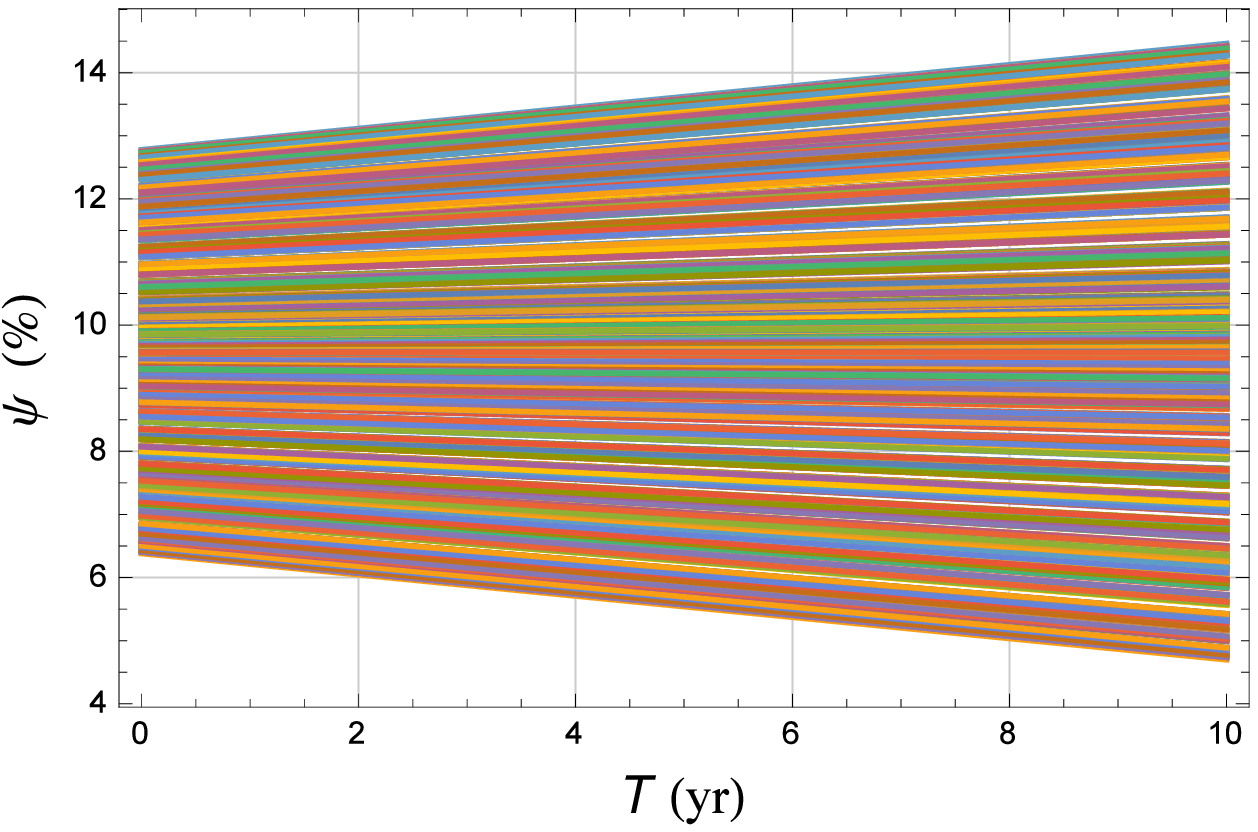}\\
\end{tabular}
}
}
\caption{Scatter over $10~\textrm{yr}$ of the per cent error $\psi$ calculated by means of \rfrs{LTvero}{J2vero} in \rfr{psi} by allowing $\Omega_0^\textrm{L},~\Omega_0^\textrm{CL}$ to vary independently of each other within a range $360~\textrm{deg}$ wide. The values $\delta{\overline{C}}_{2,0}=2.5\times 10^{-10}$ (see Table~\ref{zerotide}), and $\Delta a^\textrm{CL} = 20~\textrm{km}$, $\Delta I^\textrm{CL} = 0.5~\textrm{deg}$  were adopted for CL.}\label{preces}
\end{figure*}

A further source of potentially non-negligible systematic uncertainty is as follow.
The precessions of \rfrs{LTO}{NodoJ2}, on which \rfr{psi} is based, hold in a coordinate system whose reference $z$ axis is aligned with the Earth's spin axis. Actually, real data reductions are performed with respect to the International Celestial Reference Frame (ICRF) \citep{2010ITN....36....1P} which adopts  the mean Earth's equator at epoch J2000.0 as fundamental reference $\grf{x,~y}$ plane. In fact, $\bds{\hat{S}}$ does vary in time because  of a variety of physical phenomena among which the lunisolar torques inducing the precession of the equinoxes is a major one. Thus, in  correctly assessing the systematic bias due to $J_2$ one has to properly account for the fact that, at the time of the data reduction, $\bds{\hat{S}}$ will be displaced by a certain amount with respect to the case of \rfrs{LTO}{NodoJ2}, and adequate formulas have to be used. To this aim, the general gravitomagnetic and $J_2$-driven node precessions are \citep{2017EPJC...77..439I}
\eqi
\dot\Omega_\textrm{LT} = \rp{2GS\csc I}{c^2 a^3\ton{1-e^2}^{3/2}}\bds{\hat{S}}\bds\cdot\bds{\hat{\mathrm{m}}}\lb{LTvero}
\eqf
\eqi
\dot\Omega_{J_2} = -\rp{3}{2}\nk\ton{\rp{R}{a}}^2\rp{\csc I~J_2}{\ton{1-e^2}^2}\ton{\bds{\hat{S}}\bds\cdot\bds{\hat{\mathrm{m}}}}\ton{\bds{\hat{S}}\bds\cdot\bds{\hat{\mathrm{n}}}}.\lb{J2vero}
\eqf
\textcolor{black}{In \rfrs{LTvero}{J2vero}, $\bds{\hat{\mathrm{m}}}$ is a unit vector in the orbital plane directed transversely to the line of the nodes, while $\bds{\hat{\mathrm{n}}}$ is a unit vector perpendiculrar to the orbital plane along the orbital angular momentum.}
%\textcolor{black}{For the definition of the unit vectors $\bds{\hat{\mathrm{m}}},~\bds{\hat{\mathrm{n}}}$ in \rfrs{LTvero}{J2vero}, see Appendix~\ref{appen}.}
About $\bds{\hat{S}}$, its temporal variation\textcolor{black}{, which introduces a time-dependence in the LT and $J_2$ rates of \rfrs{LTvero}{J2vero},} can be expressed as \citep{2007CeMDA..98..155S}
\begin{align}
\alpha & \simeq 0.00 - 0.641~\textrm{deg~cty}^{-1}~\tau\lb{alfa}, \\ \nonumber \\
\delta & \simeq 90~\textrm{deg} -0.557~\textrm{deg~cty}^{-1}~\tau\lb{delta}.
\end{align}
\textcolor{black}{In \rfrs{alfa}{delta}, $\alpha,~\delta$ are the right ascension (RA) and the declination (DEC) of the Earth's spin axis, respectively, while $\tau$ is the interval in Julian centuries (of 36525 days) from the standard epoch JD 2451545.0. The Earth's spin axis can be expressed in terms of $\alpha,~\delta$ as \eqi
\bds{\hat{S}} = \grf{\cos\alpha\cos\delta,~\sin\alpha\cos\delta,~\sin\delta}.\lb{SRADEC}
\eqf
}
%\textcolor{black}{For the definition and the meaning of $\bds{\hat{S}},~\alpha,~\delta,~\tau$,  see Appendix~\ref{appen}.}
Thus, if one calculates \rfr{psi} by means of \rfrs{LTvero}{delta}, a dependence on $t$ and the initial values $\Omega_0^\textrm{L},~\Omega_0^\textrm{CL}$ of the nodes of L and CL at the starting epoch of the data reduction which may have a non-negligible impact is introduced.
\textcolor{black}{By expanding \rfr{SRADEC} according to \rfrs{alfa}{delta} and neglecting terms quadratic in $\dot\alpha,~\dot\delta$, a linear dependence on $t$ occurs. Indeed, one has
\eqi
\bds{\hat{S}} \simeq \grf{-\dot\delta~\tau,~\mathcal{O}\ton{\dot\alpha\dot\delta},~1 + \mathcal{O}\ton{\dot\delta^2}}.
\eqf
It implies a quadratic signature in the integrated node residuals giving rise to a mild parabolic residual signal. Over temporal intervals just some years long, it would likely superimpose to the linear LT trend by potentially corrupting its recovery. A similar issue occurs also for other potential sources of systematic errors (see Sections~\ref{decay}~to~\ref{decayI}).}
Figure~\ref{preces} depicts the scatter in the values of $\psi$ over $10~\textrm{yr}$ for $\delta{\overline{C}}_{2,0} = 2.5\times 10^{-10}$ (see Table~\ref{zerotide}) and $\Delta a^\textrm{CL} = 20~\textrm{km},~\Delta I^\textrm{CL} = 0.5~\textrm{deg}$ by allowing $\Omega_0^\textrm{L},~\Omega_0^\textrm{CL}$ to vary independently of each other within a $360~\textrm{deg}$ range. It can be noted that $\psi$ finally vary from $5\%$ to about $15\%$.
\section{The indirect impact of the Earth's oblateness through the secular decay of the satellites' semimajor axes}\lb{decay}
It has been known for several decades that the semimajor axis $a$ of the existing LAGEOS-type satellites experiences a secular decay $\dot a<0$ as large as \citep{1982CeMec..26..361R,Sosbook,Sosproc,2017AcAau.140..469P}
\eqi
\left|\dot a\right|\simeq 0.2-0.9~\textrm{m~yr}^{-1}
\eqf
experimentally determined with an accuracy of the order of \citep{1982CeMec..26..361R,Sosbook,Sosproc}
\eqi
\sigma_{\dot a}\simeq~0.03-0.1~\textrm{m~yr}^{-1};
\eqf
see Table~2 of \citet{2016AdSpR..57..493I}.
About LAGEOS, it should be remarked that the experimental accuracy in determining its orbital decay rate has not significantly improved over the past few decades. Indeed,  \citet{1982CeMec..26..361R} reported $\sigma_{\dot a^\textrm{L}}=0.1~\textrm{mm~d}^{-1}=0.036~\textrm{m~yr}^{-1}$, while $\sigma_{\dot a^\textrm{L}} = 0.035~\textrm{m~yr}^{-1}$ is quoted in \citet{Sosbook,Sosproc}. \textcolor{black}{Note that the quoted $\sigma_{\dot a}$ are not due to a priori uncertainties in the acceleration models used; instead, they are a posteriori errors generated in the data reduction procedure.}

Neither the pN gravitomagnetic field of the Earth nor its Newtonian oblateness $J_2$ directly affect the semimajor axis $a$ with secular perturbations. Nonetheless, as shown by \citet{2016AdSpR..57..493I}, the secular decay $\dot a<0 $ of such an orbital element has a long-term indirect effect on the node $\Omega$ through its interplay with its classical precession $\dot\Omega_{J_2}$ driven by $J_2$. The resulting change over a time span $T$ is \citep{2016AdSpR..57..493I}
\eqi
\Delta\Omega_{J_2}^{\dot a}\simeq \rp{21}{8}\ton{\rp{R}{a_0}}^2\rp{J_2\cos I}{\ton{1-e^2}^2}\ton{\rp{\nk \dot a T^2}{a_0}}+\mathcal{O}\ton{\dot a^2 T^3}\lb{OJ2a}.
\eqf
\textcolor{black}{In \rfr{OJ2a}, $a_0$ is the semimajor axis at some reference epoch assumed as initial value at the beginning of the observational time span $T$.}
To the first order in $\dot a$, \rfr{OJ2a} can be written as
\eqi
\Delta\Omega_{J_2}^{\dot a} = K_{J_2}\dot a,
\eqf
where
\eqi
K_{J_2}\doteq\derp{\Delta\Omega_{J_2}^{\dot a}}{\dot a}=\rp{21}{8}\ton{\rp{R}{a_0}}^2\rp{J_2\cos I}{\ton{1-e^2}^2}\ton{\rp{\nk  T^2}{a_0}}.
\eqf

\begin{figure*}
\centerline{
\vbox{
\begin{tabular}{c}
\epsfbox{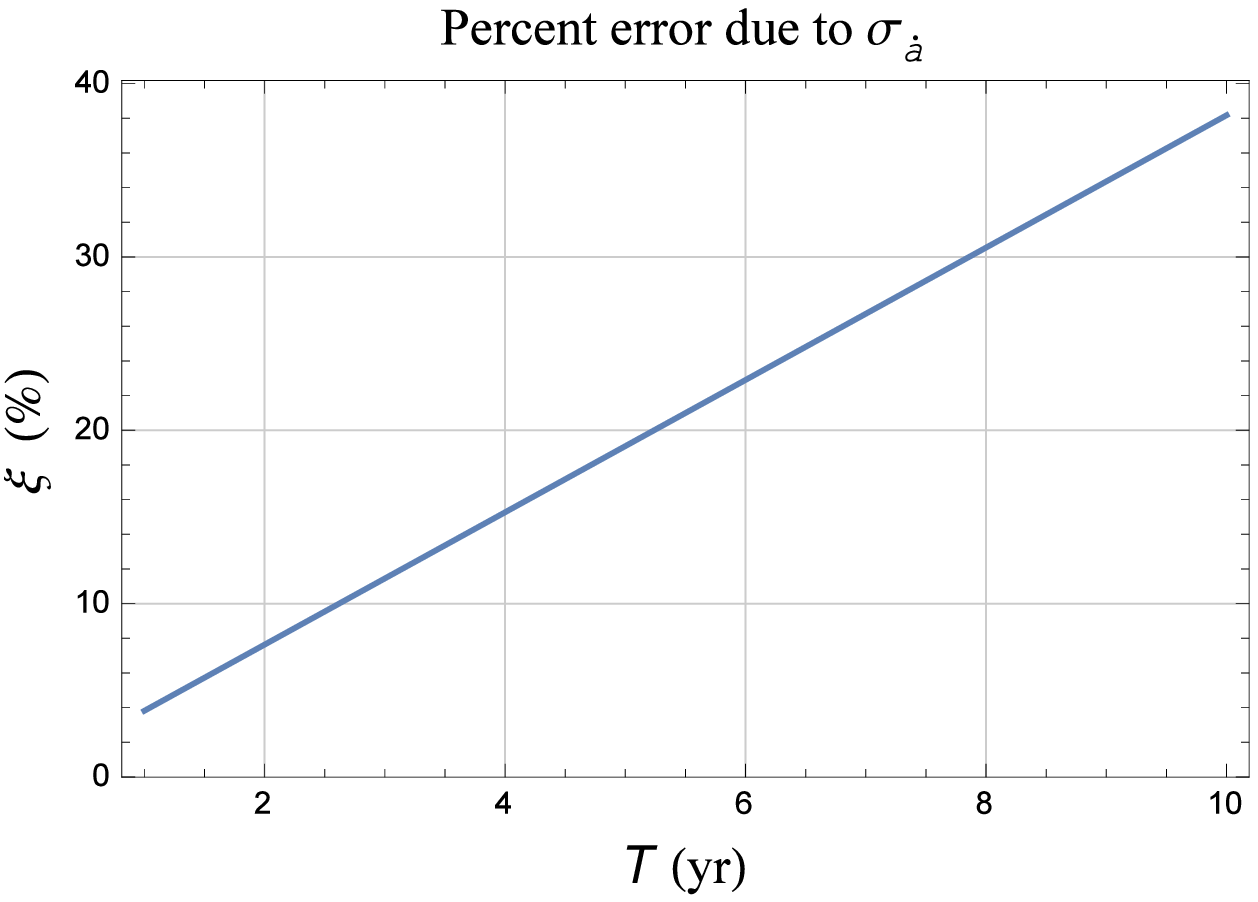}\\
\end{tabular}
}
}
\caption{Percent systematic error $\xi$ in the total $\textrm{L} + \textrm{CL}$ Lense-Thirring node shift due to the uncertainties in the secular decays of the semimajor axes of L and CL calculated as a function of the data analysis time span $T$ with \rfrs{errsma}{csi}. For CL we assumed $\Delta a^\textrm{CL} = 10~\textrm{km},~\Delta I^\textrm{CL} = 0.15~\textrm{deg},~\sigma_{\dot a^\textrm{CL}}=0.01~\textrm{m~yr}^{-1}$, while for L the value of the uncertainty in the decay of its semimajor axis adopted is $\sigma_{\dot a^\textrm{L}}=0.035~\textrm{m~yr}^{-1}$. \textcolor{black}{In line with with what has been observed  for the last decades in the case of LAGEOS, we assumed that $\sigma_{\dot a^\textrm{L}}~\sigma_{\dot a^\textrm{CL}}$ stay constant during $T$.} }\label{figura1}
\end{figure*}

It is quite plausible and reasonable to assume that also CL will finally experience such a subtle orbital decay\textcolor{black}{, although much will likely depend on the actual manufacturing of CL and its surface properties}. Thus, it is important to calculate its impact on the proposed use of the sum of the nodes of LAGEOS and CL.
It is worthwhile remarking that we will exclusively rely upon the so-far phenomenologically determined values of the secular decays of the semimajor
axes of the satellites of the LAGEOS family from real data reductions of long observational records. Stated differently, we will not try to model such an observed orbital feature in terms of some exotic or mundane physical phenomena; for several attempts towards this goal in terms of standard non-gravitational effects, see, e.g., \citet{1982CeMec..26..361R,1987JGR....92.1287R,1988JGR....9313805R,1991JGR....96..729S,2015CQGra..32o5012L,2017AcAau.140..469P}.
Since each satellite experiences its own semimajor axis secular rate, $\dot a$ must be treated as an independent variable for both satellites in calculating the associated error in the sum of their node shifts which, thus, reads
\eqi
\delta\ton{\Delta\Omega^\textrm{tot}_{\dot a}} \leq\left|K_{J_2}^\textrm{L}\right|\sigma_{\dot a^\textrm{L}} + \left|K_{J_2}^\textrm{CL}\right|\sigma_{\dot a^\textrm{CL}}.\lb{errsma}
\eqf
Note that it was assumed that the errors $\sigma_{\dot a^\textrm{L}},~\sigma_{\dot a^\textrm{CL}}$ stay constant during the data analysis. Should they change, things would be even more complicated.
In this case, contrary to the bias due to the Earth's even zonal $J_2$, the peculiar orbital configuration of CL does not provide benefits in calculating the systematic error due to $\sigma_{\dot a}$. Indeed, if $\dot a$ were, say, an external parameter common to both satellites as it occurs for $J_2$, the coefficient proportional to $\sigma_{\dot a}$ would be $\left|K_{J_2}^\textrm{L}+K_{J_2}^\textrm{CL}\right|$.
As a result, the corresponding percent error
\eqi
\xi\doteq \rp{\delta\ton{\Delta\Omega^\textrm{tot}_{\dot a}}}{\Delta\Omega_\textrm{LT}^\textrm{tot}}\lb{csi}
\eqf
in the total Lense-Thirring node shift, for given values of $\Delta a^\textrm{CL},~\Delta I^\textrm{CL}$ and $\sigma_{\dot a^\textrm{L}},~\sigma_{\dot a^\textrm{CL}}$, is linear in $T$. Figure~\ref{figura1} depicts it over a time shift 10 yr long for $\Delta a^\textrm{CL} = 10~\textrm{km},~\Delta I^\textrm{CL} = 0.15~\textrm{deg},\sigma_{\dot a^\textrm{L}}=0.035~\textrm{m~yr}^{-1}$. For CL, we hypothesize an improvement of, say, a factor of 3 in the determination of its possible orbital rate decay with respect to LAGEOS by assuming $\sigma_{\dot a^\textrm{CL}}=0.01~\textrm{m~yr}^{-1}$.
It can be noted that $\xi$ is as large as $20\%$ after 5 yr, reaching $40\%$ after 10 yr. It turns out that, even if it were $\sigma_{\dot a^\textrm{CL}}=0$, the situation would not change too much, with a maximum bias of roughly $\simeq 35\%$ after 10 yr. A breakthrough in the accuracy of the determination
of the orbital decay rate of LAGEOS by a factor of 10 would be needed to bring such source of systematic uncertainty down to the percent level; in view of what happened so far in the last decades, it does not seem plausible to occur in the foreseeable future. \textcolor{black}{At least as far as the existing LAGEOS is concerned, the experimental error $\sigma_{\dot a^\textrm{L}}$ has shown so far a very weak time dependence, staying essentially constant over the decades. }

For the sake of completeness, we also note that the impact of the uncertainty in $J_2$ in the sum of the corresponding node shifts due to $\dot a$ of \rfr{OJ2a} is negligible with respect to the added Lense-Thirring node rates.

\textcolor{black}{Finally, it is worth noting that the effect considered in this Section would act as some sort of parabolic signature in the integrated node residuals, being quadratic in time. Nonetheless, it would be premature and unjustified to argue that its mismodelled component would necessarily decouple from the LT linear signal. Indeed, the resulting parabola would be rather flat, especially over not too long temporal intervals $T$. Thus, it is likely that the recovery of the gravitomagnetic slope would be impacted by the aforementionded mismodeled competing quadratic effect. A quantitative assessment of the level of such a potential bias is beyond the scope of the present paper since it would require ad-hoc data simulations and reductions.}
%
%{NodoJ2}
\section{The indirect impact of the Earth's oblateness through the secular decay of the satellites' inclination}\lb{decayI}
Despite it seems that, at present, no experimental determinations of a possible secular decay $\dot I$ of the inclination of the LAGEOS-type satellites are publicly available in the literature, there are several \textcolor{black}{standard} physical mechanisms of non-gravitational origin which, in principle,  are able to induce such an effect. Thus, \textcolor{black}{despite the content of the present Section may be deemed as still hypothetical,} we will \textcolor{black}{prefer to} accurately consider \textcolor{black}{also} its possible indirect impact on the classical node precessions driven by the Earth's oblateness as done for the semimajor axis decay in Section~\ref{decay}.

By inserting
\eqi
I(t)\simeq I_0+\dot I t,\lb{It}
\eqf
in \rfr{NodoJ2}, the total node shift including also the part due to the secular decay of the inclination, integrated over a time span $T$, is
\eqi
\Delta\Omega_{J_2}=\int_0^T \dot\Omega_{J_2}\ton{t}dt = \rp{3\nk R^2 J_2\qua{\sin{I_0}-\sin{\ton{I_0+\dot I T}}}}{2a^2\ton{1-e^2}^2\dot I}.\lb{DO}
\eqf
\textcolor{black}{In \rfrs{It}{DO}, $I_0$ is the inclination at some reference epoch assumed as initial value at the beginning of the observational time span $T$.}
Note that, in the limit $\dot I\rightarrow 0$, \rfr{DO} reduces just to \rfr{NodoJ2}.
By  posing
\eqi
\theta\doteq \dot I T\ll 1,\lb{alph}
\eqf
as it is the case for the LAGEOS-type satellites over even multidecadal time spans $T$ (see \rfr{07L} and \rfr{04CL} below),
let us expand the trigonometric functions $\sin\alpha,~\cos\alpha$  in $\sin\ton{I_0 + \alpha}$ entering \rfr{DO} as
\begin{align}
\cos\theta &\simeq 1 - \rp{\theta^2}{2},\\ \nonumber \\
\sin\theta &\simeq \theta - \rp{\theta^3}{6}.
\end{align}
Thus,  from \rfr{DO} one can approximately obtain for the $\dot I$-induced node shift
\eqi
\Delta\Omega^{\dot I}_{J_2}\simeq \rp{\nk J_2 R^2 \dot I T^2\ton{\dot I T\cos I_0 + 3\sin I_0  }}{4 a^2\ton{1-e^2}^2}.\lb{DOapprox}
\eqf
From \rfr{DOapprox}, it turns out
\eqi
\delta\ton{\Delta\Omega^{\dot I}_{J_2}}\leq \left|\derp{\Delta\Omega^{\dot I}_{J_2}}{\dot I}\right|\sigma_{\dot I}\simeq \left|\rp{\nk J_2 R^2 T^2\ton{2\dot I T\cos I_0+3\sin I_0}}{4 a^2\ton{1-e^2}^2}\right|\sigma_{\dot I}.
\eqf
By posing
\eqi
\mathcal{I}_{J_2}\doteq \rp{\nk J_2 R^2 T^2\ton{2\dot I T\cos I_0 + 3\sin I_0}}{4 a^2\ton{1-e^2}^2},\lb{IJ2}
\eqf
the total systematic bias in the sum of the nodes is, thus,
\eqi
\delta\ton{\Delta\Omega^\textrm{tot}_{\dot I}} \leq\left|\mathcal{I}_{J_2}^\textrm{L}\right|\sigma_{\dot I^\textrm{L}} + \left|\mathcal{I}_{J_2}^\textrm{CL}\right|\sigma_{\dot I^\textrm{CL}}.\lb{errI}
\eqf

\begin{figure*}
\centerline{
\vbox{
\begin{tabular}{c}
\epsfbox{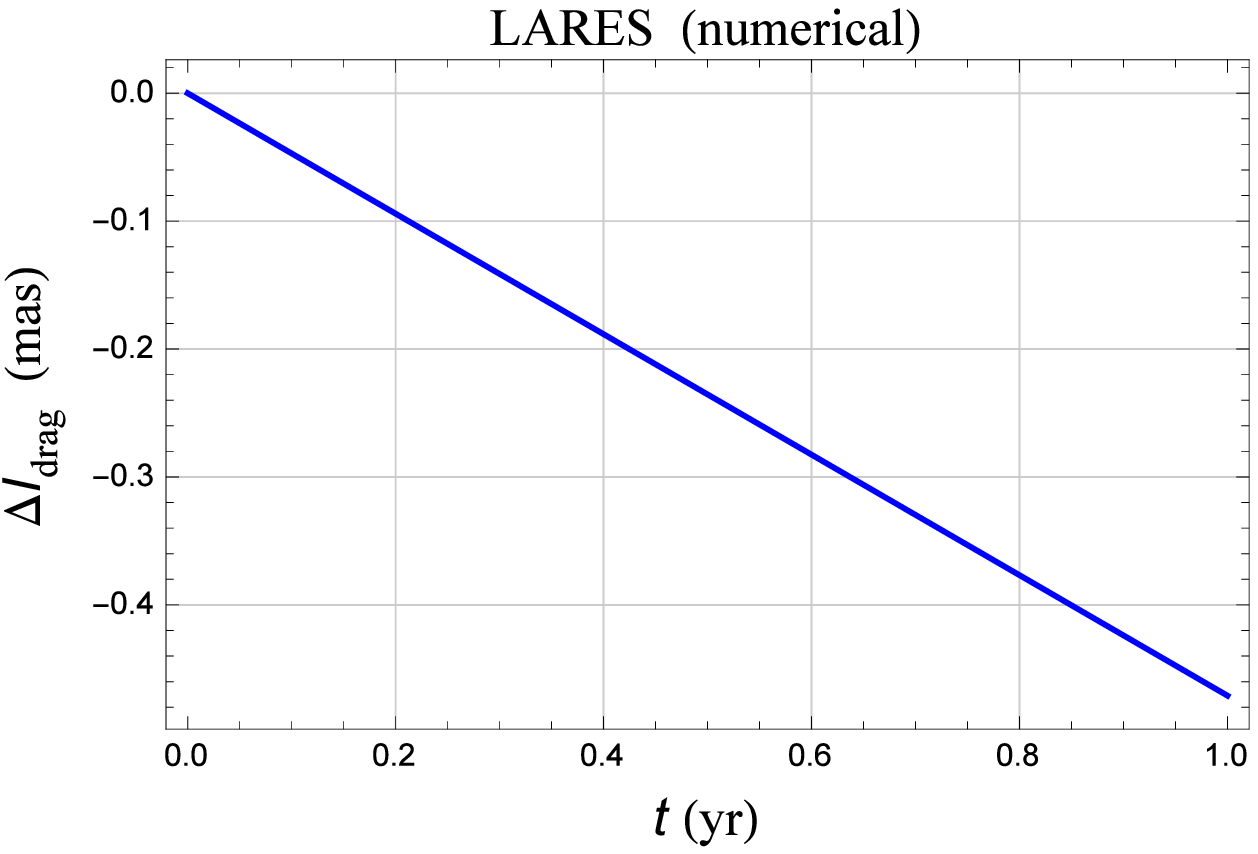}\\
\end{tabular}
}
}
\caption{Numerically produced time series of the inclination shift $\Delta I_\textrm{drag}$, in mas, of LARES over a time span $T = 1~\textrm{yr}$  as a result of the difference of two numerical integrations of its equations of motion in Cartesian coordinates with and without the drag acceleration, calculated with the values of the atmospheric and satellite parameters from \citet{2017AcAau.140..469P}, sharing the same initial conditions for August 6, 2012 retrieved on the Internet at https://www.calsky.com/. A secular rate of
$\dot I^\textrm{LR}_\textrm{drag} = -0.5~\textrm{mas~yr}^{-1}$ is neatly apparent, in agreement with the analytical result in \citet{2010AcPPB..41.4753I}.}\label{figura2}
\end{figure*}

\begin{figure*}
\centerline{
\vbox{
\begin{tabular}{c}
\epsfbox{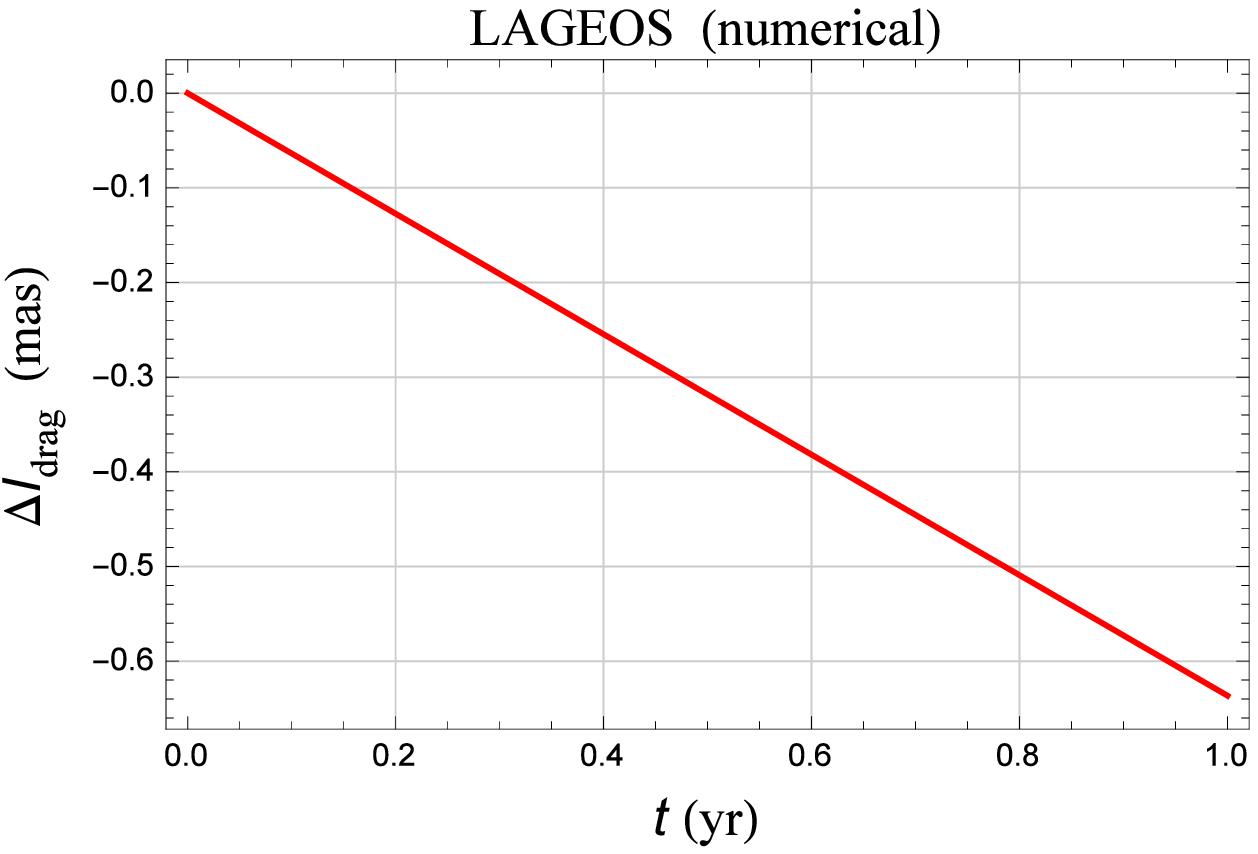}\\
\end{tabular}
}
}
\caption{Numerically produced time series of the inclination shift $\Delta I_\textrm{drag}$, in mas, of LAGEOS over a time span $T = 1~\textrm{yr}$  as a result of the difference of two numerical integrations of its equations of motion in Cartesian coordinates with and without the drag acceleration, calculated with the values of the neutral and charged drag parameters of \rfrs{Hdrag}{AtoM}, sharing the same initial conditions provided in a personal communication to the author by a colleague. A secular rate of
$\dot I^\textrm{L}_\textrm{drag} = -0.7~\textrm{mas~yr}^{-1}$ is neatly apparent.}\label{figura3}
\end{figure*}

Not only the experimental accuracy $\sigma_{\dot I}$ in measuring a possible decay of the orbital plane, but also its nominal value $\dot I$  enter \rfr{errI} through \rfr{IJ2}. Thus, at least in principle, it is important to try to give a plausible order of magnitude of such a subtle phenomenon.
Below, we will consider only the effect of the neutral and charged drag, whose disturbing acceleration can be modeled as
\citep{1987soaa.book.....K,1987ahl..book.....M}
\eqi
{\bds A}_\textrm{drag} = -\rp{1}{2}C_\textrm{D}\Sigma\rho V\bds{V}.\lb{Adrag}
\eqf
\textcolor{black}{In \rfr{Adrag}, $C_\textrm{D},~\Sigma,~\rho,~\bds{V}$ are the satellite's drag coefficient and area-to-mass ratio, the atmospheric density at the satellite's height, and the satellite's velocity with respect to the Earth's atmosphere, respectively.}
It can be shown that, among other things, \rfr{Adrag} also induces a secular variation of the inclination $I$ given by \citep{1987ahl..book.....M,2010AcPPB..41.4753I}
\eqi
\dert{I}{t} = -\rp{1}{4}C_\textrm{D}\Sigma\rho\omega_\textrm{atm}a\sin I.\lb{Idrag}
\eqf
\textcolor{black}{In \rfr{Idrag}, $\omega_\textrm{atm}$ is the angular velocity of Earth's atmosphere.}
We will, first, confirm the validity of \rfr{Idrag} in the case of LARES by comparing its predicted theoretical rate with a numerical integration of its equations of motion.
For such a satellite it is \citep{2017AcAau.140..469P}
\begin{align}
C_\textrm{D} & = 3.5\lb{cd}, \\ \nonumber \\
\Sigma & = 2.69\times 10^{-4}~\textrm{m}^2~\textrm{kg}^{-1}, \\ \nonumber\\
\rho_\textrm{neutral} & = 5.9\times 10^{-16}~\textrm{kg}~\textrm{m}^{-3}, \\ \nonumber \\
\omega_\textrm{atm} &\simeq\omega_\oplus  = 7.29\times 10^{-5}~\textrm{s}^{-1}\lb{om},
\end{align}
so that \rfr{Idrag} allows to predict a nominal decay of the order of \citep{2010AcPPB..41.4753I}
\eqi
\dot I^\textrm{LR}_\textrm{drag} = -0.5~\textrm{mas}~\textrm{yr}^{-1}.\lb{IdragLR}
\eqf
\textcolor{black}{In \rfr{om}, $\omega_\oplus$ is the angular velocity of Earth.}
The analytical result of \rfr{IdragLR} is supported by a numerical integration of the equations of motion of LARES in rectangular Cartesian coordinates with and without \rfr{Adrag}. Figure~\ref{figura2} displays the resulting time series for the drag-induced inclination shift $\Delta I_\textrm{drag}(t)$ of LARES over a time span 1 yr long; a negative trend with just the same size of \rfr{IdragLR} is clearly apparent. It can be shown that our approach is also able to reproduce, both analytically and numerically,  the observed secular decay of the semimajor axis of LARES, in agreement with the finding by \citet{2017AcAau.140..469P} who found that the neutral atmospheric drag is able to explain almost entirely such a phenomenon.
Thus, confident of our strategy, we can, now, safely apply it to LAGEOS and CL.
At the altitude of 5900 km, the total neutral atmospheric density experienced by LAGEOS is of the order of
\eqi
\rho_\textrm{neutral}^\textrm{L}\simeq 8.4\times 10^{-18}~\textrm{kg}~\textrm{m}^{-3}\lb{Hdrag}
\eqf
due to neutral hydrogen H, corresponding to a hydrogen number density of $5\times 10^9~\textrm{m}^{-3}$ \citep{1990JGR....95.4881R}.
As far as the charged particles in the plasmasphere are concerned,
\eqi
\rho_{\textrm{H}^+}^\textrm{L}\simeq 5.0\times 10^{-18}~\textrm{kg}~\textrm{m}^{-3}\lb{pt}
\eqf
comes from protons H$^+$ for a proton number density of $3\times 10^9~\textrm{m}^{-3}$ \citep{1990JGR....95.4881R},
\eqi
\rho_{\textrm{He}^+}^\textrm{L}\simeq 7.2\times 10^{-18}~\textrm{kg}~\textrm{m}^{-3}
\eqf
refers to Helium ions He$^+$, while
\eqi
\rho_{\textrm{O}^+}^\textrm{L}\simeq 4.6\times 10^{-17}~\textrm{kg}~\textrm{m}^{-3}\lb{oxy}
\eqf
is  attributable to Oxygen ions O$^+$.
However, as noted in \citet{1990JGR....95.4881R}, \rfrs{pt}{oxy} have to be divided by a factor of 2 since LAGEOS spends just half its time in the plasmasphere. Thus, the total charged density actually felt by LAGEOS is about
\eqi
\rho_\textrm{ch}^\textrm{L}\simeq 2.9\times 10^{-17}~\textrm{kg}~\textrm{m}^{-3}.
\eqf
In calculating the acceleration due to charged drag with \rfr{Adrag}, the charged drag coefficient of LAGEOS can be as large as  \citep{1987ahl..book.....M}
\eqi
C_\textrm{D}^\textrm{ch}\simeq 20,\lb{CDch}
\eqf
while its area-to-mass ratio amounts to
\eqi
\Sigma = 7\times 10^{-4}~\textrm{m}^2~\textrm{kg}^{-1}.\lb{AtoM}
\eqf
Thus, both \rfr{Idrag} and a numerical integration of the equations of motion, whose outcome is displayed in Figure~\ref{figura3}, return \textcolor{black}{a secular} decay of the orbital plane of LAGEOS due to neutral and charged atmospheric drag of the order of
\eqi
\dot I^\textrm{L}_\textrm{drag} \simeq -0.7~\textrm{mas}~\textrm{yr}^{-1}.\lb{07L}
\eqf
In the case of CL, since from the satellite's physical parameteres released by \citet{2017EPJP..132..336C} it can be inferred
\eqi
\Sigma = 4\times 10^{-4}~\textrm{m}^2~\textrm{kg}^{-1},
\eqf
the drag-induced inclination decay would amount to
\eqi
\dot I^\textrm{CL}_\textrm{drag} \simeq -0.4~\textrm{mas}~\textrm{yr}^{-1}\lb{04CL}
\eqf
by assuming the same value of \rfr{CDch} for its $C_\textrm{D}^\textrm{ch}$. Incidentally, \rfr{07L} and \rfr{04CL} confirm the validity of the assumption made in \rfr{alph}.
Also thermal effects connected with the orientation and magnitude of the satellite's spin axis \citep{1987JGR....92.1287R} can induce a secular decay on $I$ \citep{2002P&SS...50.1067L}; thus, the actual overall size of such a phenomenon may be larger than \rfr{07L} and \rfr{04CL}.

\begin{figure*}
\centerline{
\vbox{
\begin{tabular}{c}
\epsfbox{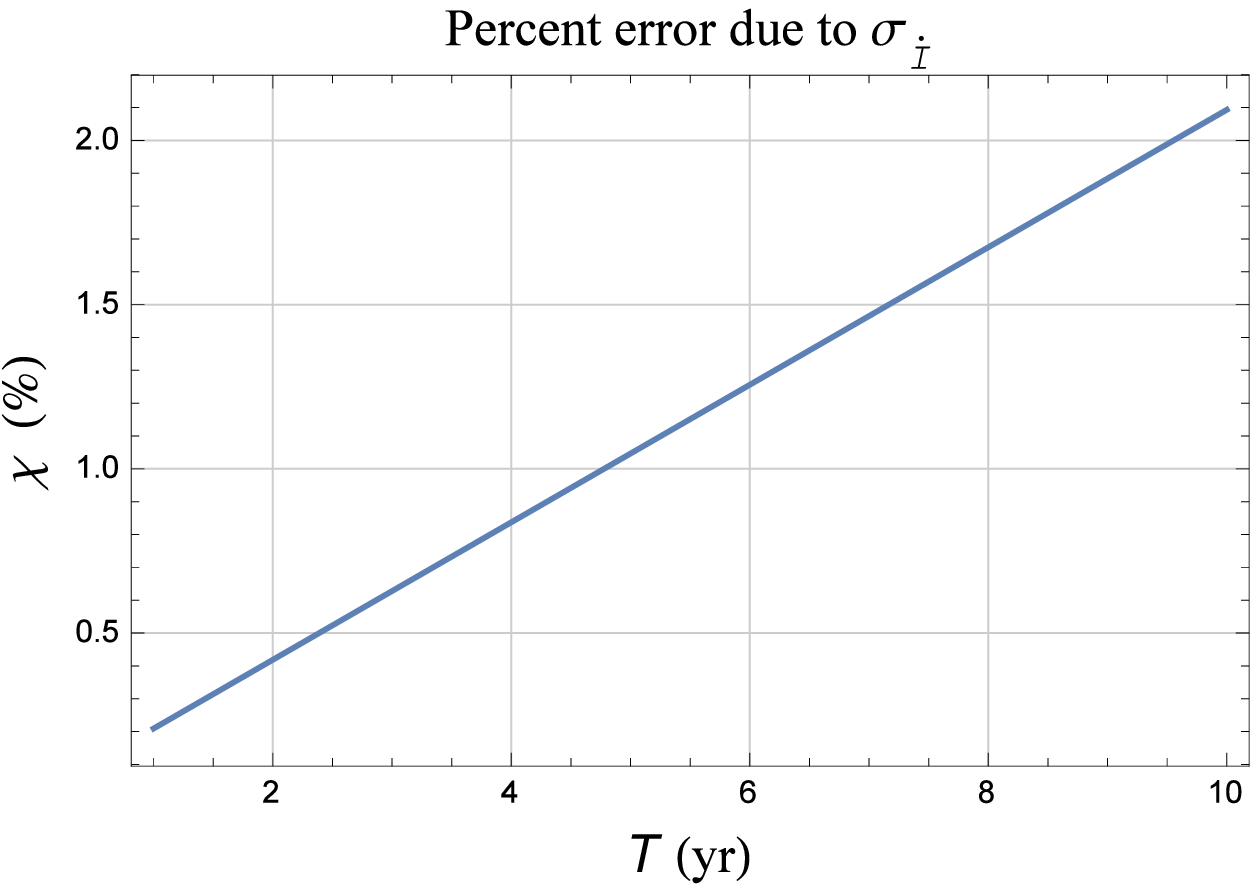}\\
\end{tabular}
}
}
\caption{Percent systematic error $\chi$ in the total $\textrm{L} + \textrm{CL}$ Lense-Thirring node shift due to the uncertainties in the presumed secular decay of the inclinations of L and CL calculated as a function of the data analysis time span $T$ with \rfr{errI} into \rfr{chi}. For CL we assumed $\Delta a^\textrm{CL} = 10~\textrm{km},~\Delta I^\textrm{CL} = 0.15~\textrm{deg},~\dot I^\textrm{L} = -0.7~\textrm{mas~yr}^{-1},~\dot I^\textrm{CL} = -0.4~\textrm{mas~yr}^{-1},~\sigma_{\dot I^\textrm{L/CL}}=0.03~\textrm{mas~yr}^{-1}$. By raising $\sigma_{\dot I}$ up to, say, $0.2~\textrm{mas~yr}^{-1}$ would bring $\chi$ to about $14\%$ after 10 yr. }\label{figura4}
\end{figure*}

For some reasons, only $\dot a$ has been experimentally investigated so far by satellite geodesists; it is expected that, sooner or later,  they will look also at the inclination by releasing a best estimate for the predicted inclination decay $\dot I$ along with its associated error $\sigma_{\dot I}$. For the sake of convenience, we may tentatively assume
\eqi
\sigma_{\dot I}\simeq \ton{5-10\%}\dot I = 0.035-0.07~\textrm{mas}~\textrm{yr}^{-1};
\eqf
it is the same percent error of the actually measured $\dot a$. \textcolor{black}{The present assumption that the expected secular decay of the inclination will be measured with the same fractional accuracy as the decay of the semimajor axis should be regarded just as a, hopefully, plausible guess in view of the lingering lacking of actual experimental determinations of it.} Note that \citet{2009SSRv..148...71C} claim to be able to determine the inclination of LAGEOS within an accuracy of just $30~\mu\textrm{as}=0.03~\textrm{mas}$.
\textcolor{black}{About the overall temporal pattern of the bias due to such a potential source of systematic error, the same considerations as in Sections~\ref{geopot}~to~\ref{decay} about the temporal variation of the Earth's spin axis and the cross correlation between $J_2$ and $\dot a$ hold.} Figure~\ref{figura4} depicts
\eqi
\chi\doteq \rp{\delta\ton{\Delta\Omega^\textrm{tot}_{\dot I}}}{\Delta\Omega_\textrm{LT}^\textrm{tot}}\lb{chi}
\eqf
for given values of $\Delta a^\textrm{CL},~\Delta I^\textrm{CL},~\dot I^\textrm{L},~\dot I^\textrm{CL},~\sigma_{\dot I^\textrm{L}},~\sigma_{\dot I^\textrm{CL}}$ over a time shift 10 yr long. For the orbital shifts of CL we assumed $\Delta a^\textrm{CL} = 10~\textrm{km},~\Delta I^\textrm{CL} = 0.15~\textrm{deg}$. As far as the inclination decays and their putative errors, we took $\sigma_{\dot I^\textrm{L/CL}}=0.03~\textrm{mas~yr}^{-1}$ along with \rfr{07L} and \rfr{04CL}. It turns out that the nominal value $\dot I$ of the expected inclination decay does not has a relevant impact on the total bias $\chi$; even by rescaling $\dot I$ by a factor of 10 for both satellites do not alter it. Thus, accurately predicting all the possible contributions to $\dot I$ does not appear, actually, so important in the present context. On the contrary, $\sigma_{\dot I}$ is of crucial importance. Indeed, while for their previously listed values the total bias can reach $2\%$ of the summed LT shifts, as shown by Figure~\ref{figura4}, by increasing them up to, say, $\sigma_{\dot I} = 0.2~\textrm{mas~yr}^{-1}$ would push $\chi$ up to to $14\%$.
\section{Summary and conclusions}\lb{finale}
According to Ciufolini and coworkers, the sum of the precessions of the nodes of the existing LAGEOS satellite and the future CiufoLares (also known-in a more anodyne fashion-as LARES 2), to be launched in 2019 with a VEGA C rocket, should provide us with a $\simeq 0.2\%$ test of frame-dragging in the field of the Earth whose imperfectly known even zonal harmonics should contribute just $0.1\%$ the total error budget.
Actually, such an ambitious goal seems difficult to be reached because of the direct and indirect impact of the competing classical node precessions mainly induced by the first even zonal of the geopotential.

On the one hand, according to some of the latest GRACE/GOCE-based global gravity field solutions released by various international institutions in 2016-2017, the lingering scatter of their determinations of such a fundamental geophysical parameter would directly affect the sum of the nodes with an uncancelled mismodelled total precession which, in some cases, may reach even several dozens percentage points of the total Lense-Thirring effect for departures of the orbital elements of CiufoLares as little as $20~\textrm{km}$ and $0.5~\textrm{deg}$. Other more favorable scenarios point towards uncertainties of the order of $\simeq 3-10\%$. A further source of systematic uncertainty is given by the dependence on the initial values of the satellites' nodes introduced by the unavoidable displacement of the Earth's spin axis with respect to its direction at the reference epoch J2000.0 when the data reduction will be finally performed in the next years by using the ICRF. The insertion of the new SLR target in an orbit as closest as possible to the ideal butterfly configuration with LAGEOS will be crucial for effectively controlling such an insidious source of systematic error in view of the persisting difficulties in effectively constraining, at least to the desired level of accuracy for a successfull test of the Lense-Thirring effect, the first even zonal of the geopotential.

On the other hand, the classical node precession due to the oblateness of the Earth exerts a further, even subtler, but not less \textcolor{black}{potentially} insidious, aliasing effect on the relativistic signature of interest in an indirect way through its coupling with the secular decay which has been phenomenologically measured so far for the semimajor axis of all the existing members of the LAGEOS family. Indeed, it turned out that the associated bias on the sum of the nodes is unaffected by the peculiar orbital geometry of CiufoLares, being impacted crucially by the accuracy in determining the satellites' orbital decay. Unfortunately, in the case of LAGEOS, after more than 30 yr, no substantial improvements have occurred so far in improving our knowledge of the rate of decreasing of its semimajor axis. Even by assuming that the forthcoming satellite will experience a more accurately known analogous effect, the resulting systematic uncertainty may reach $\simeq 20-40\%$ of the Lense-Thirring signal after $5-10$ yr. An analogous indirect bias should occur due to the expected secular decay of the inclination of the orbital planes, although no experimental records for such an effect on the LAGEOS-type spacecraft are yet available in the literature. \textcolor{black}{Thus, such a potential further source of systematic error should be currently deemed as hypothetical, although plausible and well rooted in estabilished standard non-gravitational physics}. Depending on its actual experimental accuracy, its impact on the combined Lense-Thirring signature may be in the range $\simeq 2-14\%$ after 10 yr. We note that the indirect effects of the Earth's rotation pole position, and the secular decays of the semimajor axis and the inclination  would affect the time series of the integrated node residuals with a quadratic temporal signature. This does not necessarily mean that it would neatly decouple from the linear Lense-Thirring effect since the residual parabolic curve would be likely rather smooth and poorly distinguishable from a secular trend, especially over relatively short observational time spans. In order to quantitatively assess how much the recovery of the gravitomagnetic slope would be impacted by the aforementionded mismodeled competing quadratic effects, a full simulation of the time series of the sum of the nodes along with a best fit with a parabolic curve would be required; it is beyond the scopes of the present preliminary analysis.

In conclusion, the desired goal of a $\simeq 0.1\%$ test of the gravitomagnetic orbital precessions with LAGEOS and CiufoLares seems, at least at present, rather unlikely to be met. In addition to a very strict adherence of the actual orbit of the new spacecraft to its ideal one, a breakthrough of one order of magnitude in determining both the terrestrial first even zonal harmonic and the secular decrease of the semimajor axis of LAGEOS (and, in perspectives, of CiufoLares as well) would be required to yield a test of some percent by using the sum of the nodes.
\begin{appendices}
\section{Notations and definitions}\lb{appen}
Here, some basic notations and definitions used in the text are presented \citep{1991ercm.book.....B,Nobilibook87,2003ASSL..293.....B}
\begin{description}
\item[] $G:$ Newtonian constant of gravitation
\item[] $c:$ speed of light in vacuum
\item[] $M:$ mass of Earth
\item[] $S:$ magnitude of the angular momentum of Earth
\item[] $\bds{\hat{S}}\doteq\grf{\cos\alpha\cos\delta,~\sin\alpha\cos\delta,~\sin\delta}:$ spin axis of Earth
\item[] $\alpha:$ right ascension (RA) of Earth's north pole of rotation
\item[] $\delta:$ declination (DEC) of Earth's north pole of rotation
\item[] $\tau:$ interval in Julian centuries (of 36525 days) from the standard epoch JD 2451545.0, i.e. 2000 January 1 12 hours TDB
\item[] $R:$ equatorial radius of Earth
\item[] ${\overline{C}}_{\ell,m}:$ fully normalized Stokes coefficient of degree $\ell$ and order $m$ of the  multipolar expansion of Earth's gravitational potential
\item[] $J_\ell=-\sqrt{2\ell+1}~{\overline{C}}_{\ell,0}:$ zonal harmonic coefficient of degree $\ell$ of the multipolar expansion of Earth's gravitational potential
\item[] ${\bds{\omega}}_\oplus:$ angular velocity of Earth
\item[] ${\bds{\omega}}_\textrm{atm}:$ angular velocity of Earth's atmosphere
%\item[] $\bds r:$ satellite's position vector with respect to Earth
%\item[] ${\bds V}_\textrm{atm} = {\bds{\omega}}_\textrm{atm}\bds\times\bds{r}:$  linear velocity of Earth's atmosphere at the satellite's position
\item[] $a:$  semimajor axis of the satellite's orbit
\item[] $a_0:$  semimajor axis of the satellite's orbit at some reference epoch
\item[] $\dot a:$ nominal estimated value of the secular decay of the satellite's orbit
\item[] $\sigma_{\dot a}:$ error of the estimated value of the secular decay of the satellite's orbit
\item[] $\nk \doteq \sqrt{GM a^{-3}}:$  Keplerian mean motion of the satellite's orbit
\item[] $e:$  eccentricity of the satellite's orbit
\item[] $I:$  inclination of the orbital plane of the satellite's orbit to the primary's equator
\item[] $I_0:$  inclination at some reference epoch
\item[] $\dot I:$ nominal estimated value of the secular decay of the satellite's inclination
\item[] $\sigma_{\dot I}:$ error of the estimated value of the secular decay of the satellite's inclination
\item[] $\Omega:$  longitude of the ascending node  of the satellite's orbit
\item[] $\dot\Omega_X:$ secular node precession induced by the post-Keplerian dynamical effect $X$
\item[] $\dot\Omega_{.2}\doteq\partial\dot\Omega_{J_2}/\partial J_2:$ coefficient of the node precession due to $J_2$
\item[] $\bds{\hat{\mathrm{m}}}\doteq\grf{-\cI\sO,~\cI\cO,~\sI}:$ unit vector directed transversely to the line of the nodes in the orbital plane
\item[] $\bds{\hat{\mathrm{n}}}\doteq\grf{\sI\sO,~-\sI\cO,~ \cI}:$ unit vector of the orbital angular momentum
\item[] $T:$ temporal interval of data analysis
\item[] $C_\textrm{D}:$ satellite's drag coefficient
\item[] $\Sigma:$ satellite's cross sectional area-to-mass ratio
\item[] $\rho:$ atmospheric density at the satellite's height
%\item[] $\bds v:$ satellite's relative velocity vector with respect to the primary
\item[] $\bds V: $ satellite's velocity with respect to the Earth's atmosphere
%=\bds v-{\bds V}_\textrm{atm}
\end{description}
\section{Tables and Figures}\lb{tavole}
\begin{table*}
\caption{Relevant orbital parameters $a,~e,~I$, LT and $J_2$-driven node precessions $\dot\Omega_\textrm{LT},~\dot\Omega_{J_2}$ of the existing satellites of the LAGEOS family and of CL along with the orbital offsets for the latter quoted in \citet{2017EPJP..132..336C}. The values for the classical and relativistic node precessions of CL are calculated for its nominal orbital configuration.
%The secular decays of $a,~I$ were neglected.
}\lb{tavola0}
\begin{center}
\begin{tabular}{|l|l|l|l|l|l|}
  \hline
 Satellite  & $a$ (km) & $e$ & $I$ (deg) & $\dot\Omega_\textrm{LT}$ (mas yr$^{-1}$) & $\dot\Omega_{J_2}$ (mas yr$^{-1}$)\\
\hline
LAGEOS (L) & $12270$ & $0.0045$ & $109.84$ & $30.7$ & $-4.5\times 10^8$\\
LAGEOS II (L II) & $12163$ & $0.0135$ & $52.64$ & $31.5$ & $8.3\times 10^8$\\
LARES (LR) & $7828$ & $0.0008$ & $69.5$ & $118.1$ & $-2.24\times 10^9$\\
CiufoLares (CL) & $12270\pm 20$ & $\simeq 0$ & $70.16\pm 0.15$ & $30.7$ & $4.5\times 10^8$ \\
  \hline
\end{tabular}
\end{center}
\end{table*}
\begin{table*}
\caption{Strictly upper triangular matrix: absolute values $\Delta {\overline{C}}_{2,0}= \left|{\overline{C}}_{2,0}^{(i)}-{\overline{C}}_{2,0}^{(j)}\right|$ of the differences between the estimated normalized Stokes coefficients ${\overline{C}}_{2,0}$ of degree $\ell=2$ and order $m=0$ for the recent GRACE/GOCE-based global gravity field solutions (tide system: zero-tide) $i,~j=$Tongji-Grace02s \textcolor{black}{\citep{2016JGeod..90..503C}}, ITU$\_$GRACE16 \textcolor{black}{\citep{ITU16}}, HUST-Grace2016s \textcolor{black}{\citep{2017AdSpR..60..597Z}}, XGM2016 \textcolor{black}{\citep{2018JGeod..92..443P}} retrieved from  the section Static Models of the WEB page of the International Center for Global Earth Models (ICGEM) at http://icgem.gfz-potsdam.de/tom$\_$longtime. Strictly lower triangular matrix: maximum values of the percent error $\psi$ computed from \rfr{psi} by assuming the figures in the strictly upper triangular matrix for the uncertainty in ${\overline{C}}_{2,0}$. The semimajor axis $a$ and the inclination $I$ of CL were allowed to vary within a range of $a^\textrm{CL}=12270\pm 20~\textrm{km},~I^\textrm{CL}=70.16\pm 0.5~\textrm{deg}$, respectively. Cfr. with Figure~\ref{figzerotide}.}\lb{zerotide}
\begin{center}
\begin{tabular}{|l|l l l l|}
  \hline
   & \normalsize{Tongji-Grace02s} & \normalsize{ITU$\_$GRACE16} & \normalsize{HUST-Grace2016s} & \normalsize{XGM2016}\\
\hline
\normalsize{Tongji-Grace02s} & $-$ & $1.9\times 10^{-10}$ & $6\times 10^{-11}$ & $1.2\times 10^{-10}$ \\
\normalsize{ITU$\_$GRACE16} & $8\%$ & $-$ & $\mathbf{2.5\times 10^{-10}}$ \cellcolor[gray] {.8} & $6\times 10^{-11}$ \\
\normalsize{HUST-Grace2016s} & $2.5\%$ & $\mathbf{10\%}$ \cellcolor[gray] {.8}  & $-$ & $1.8\times 10^{-10}$ \\
\normalsize{XGM2016} & $5\%$ & $3\%$ & $8\%$ & $-$ \\
\hline
\end{tabular}
\end{center}
\end{table*}
\begin{table*}
\caption{Strictly upper triangular matrix: absolute values $\Delta {\overline{C}}_{2,0}= \left|{\overline{C}}_{2,0}^{(i)}-{\overline{C}}_{2,0}^{(j)}\right|$ of the differences between the estimated normalized Stokes coefficients ${\overline{C}}_{2,0}$ of degree $\ell=2$ and order $m=0$ for the recent GOCE-based global gravity field solutions (tide system: tide-free) $i,~j=$NULP-02s \textcolor{black}{\citep{Nulp02}}, GO$\_$CONS$\_$GCF$\_$2$\_$SPW$\_$R5 \textcolor{black}{\citep{gocons5}}, IGGT$\_$R1 \textcolor{black}{\citep{2018JGeod..92..561L}}, IfE$\_$GOCE05s \textcolor{black}{\citep{IFE05}} retrieved from  the section Static Models of the WEB page of the International Center for Global Earth Models (ICGEM) at http://icgem.gfz-potsdam.de/tom$\_$longtime. Strictly lower triangular matrix: maximum values of the percent error $\psi$ computed from \rfr{psi} by assuming the figures in the strictly upper triangular matrix for the uncertainty in ${\overline{C}}_{2,0}$. The semimajor axis $a$ and the inclination $I$ of CL were allowed to vary within a range of $a^\textrm{CL}=12270\pm 20~\textrm{km},~I^\textrm{CL}=70.16\pm 0.5~\textrm{deg}$, respectively. Cfr. with Figure~\ref{figtidefree}.}\lb{tidefree}
\begin{center}
\begin{tabular}{|l|l l l l|}
  \hline
   & \normalsize{NULP-02s} & \normalsize{GO$\_$CONS$\_$GCF$\_$2$\_$SPW$\_$R5} & \normalsize{IGGT$\_$R1} & \normalsize{IfE$\_$GOCE05s}\\
\hline
\normalsize{NULP-02s} & $-$ & $9\times 10^{-11}$ & $2.6\times 10^{-10}$ & $1.7\times 10^{-9}$ \\
\normalsize{GO$\_$CONS$\_$GCF$\_$2$\_$SPW$\_$R5} & $4\%$ & $-$ & $1.7\times 10^{-10}$  & $1.8\times 10^{-9}$ \\
\normalsize{IGGT$\_$R1} & $12\%$ & $7\%$   & $-$ & $\mathbf{1.9\times 10^{-9}}$ \cellcolor[gray] {.8}\\
\normalsize{IfE$\_$GOCE05s} & $70\%$ & $80\%$ & $\mathbf{85\%}$ \cellcolor[gray] {.8}& $-$ \\
  \hline
\end{tabular}
\end{center}
\end{table*}
\end{appendices}
\bibliography{Gclockbib,semimabib}{}

%-----------------------------------------

\end{document}